\documentclass[english,aps,preprint, superscriptaddress]{revtex4-2}
\usepackage[T1]{fontenc}
\usepackage[latin9]{inputenc}
\usepackage{float}
\usepackage{amstext}
\usepackage{graphicx}
\usepackage{hyperref}
\usepackage{amsmath}

\makeatletter

\usepackage{slashed}
\usepackage{babel}

\usepackage{babel}
\begin{document}
\title{Revisiting the two-body problem in Yukawa gravity and in a gravitational extension of the Buckingham potential}
\author{E. A. Gallegos} 
\email{egallegoscollado@gmail.com}
\affiliation{Instituto de Ci\^{e}ncia e Tecnologia, Universidade Federal de Alfenas,Campus Avan\c{c}ado de Po\c{c}os de Caldas, 37701-970, MG, Brazil}
\author{R. Perca Gonzales}
\email{rperca@unsa.edu.pe}
\affiliation{Escuela Profesional de F\'{i}sica, Facultad de Ciencias Naturales y Formales, Universidad Nacional de San Agust\'{i}n de Arequipa (UNSA), Av. Independencia S/N, Arequipa, Per\'{u}}
\author{J. F. Jesus}
\email{jf.jesus@unesp.br}
\affiliation{Departamento de Ci\^encias e Tecnologia, Instituto de Ci\^encias e Engenharia, Universidade Estadual Paulista (UNESP) - R. Geraldo Alckmin, 519, 18409-010, Itapeva, SP, Brazil}
\affiliation{Departamento de F\'isica, Faculdade de Engenharia e Ci\^encias, Universidade Estadual Paulista (UNESP) - Av. Dr. Ariberto Pereira da Cunha 333, 12516-410, Guaratinguet\'a, SP, Brazil}

\begin{abstract}
We revisit the Keplerian-like parametrization of the two-body problem in Yukawa gravity studied in the literature. Some inconsistencies, which spoil Bertrand's theorem, observed in the $\eta$ parametrization of the true anomaly $\theta$ and in the formulae for the pericenter's advance are resolved. Moreover, inspired in this kind of study, we couple the Buckingham potential, a variation of the Lennard-Jones intermolecular potential, with the gravitational
Newtonian potential and find a Keplerian-like parametrization for the solution of the two-body problem in this sort of gravity. The
outcomes for the advance of the pericenter in both types of gravity are corroborated by using the Landau and Lifshitz's method. We also tested the expressions thus obtained against Solar System and S2 star data. The result for both models is that while some deviation from general relativity (GR) is allowed, GR cannot be discarded by the current analysis.
\end{abstract}
\maketitle

\section{INTRODUCTION\protect\label{sec:Sec1}}

Hitherto, there is not a thorough understanding, based in first principles, of the two dark components, i.e., dark energy and cold dark matter
(CDM), of the most accepted cosmological model of the universe: the $\Lambda\text{CDM}$ paradigm \citep{Springeletal(2006)}. The observed  ``nearly'' constant dark energy density that suffuses the universe is represented by the cosmological constant $\Lambda$, introduced in the general theory of relativity for the first time by Einstein and removed from it later on by himself. It is important to mention, nevertheless, that the presence of the cosmological constant $\Lambda$ is mandatory in an eventual quantization of gravity (at least as an effective field theory, since this is nonrenormalizable) and this also allows us to explain, when adding to Einstein's field equation, the observed accelerating expansion of the universe. On the other hand, dark matter, an \textsl{ad hoc} assumption, is conceived to account for some gravitational effects on galaxies and larger structures, e. g. the Doppler shifts of stellar motions near the plane of the Milky Way \citep{Oort(1932)} and the anomalous motions of the galaxies in the Coma Cluster \citep{Zwicky(1933)}, among others, by assuming the correctness of general relativity (GR). This kind of invisible matter, by definition, does not interact electromagnetically and, up to now, no dark matter particle has been observed and identified. To avoid or explain the hypothetical dark components, in particular, dark matter, alternative theories of gravity were proposed by altering the general theory of relativity such as modified Newtonian dynamics \citep{Milgrom(1983),Sanders(1984)}, extended theories of gravity [$f\left(R\right)$ gravity, for instance] \citep{Buchdahl(1970)} or entropic gravity \citep{Verlinde(2011)}. 

Recently, Yukawa-like gravity theories have attracted considerable attention mainly for two reasons: firstly, some studies \citep{Jusufietal(2023),Gonzalesetal(2023),Jovanovicetal(2023)} highlight that they would have potential to account for both dark matter and dark energy at galactic and cosmological scales, respectively; and secondly, since these theories are massive gravity ones (see \citep{FierzPauli(1939),deRahametal(2017)} and references therein), the frequency dependence of velocity of the observed gravitational waves \citep{deRahametal(2017),LIGO-VIRGO(2016)} might be explained through the small but nonzero mass of the graviton (the spin-2 carrier of the gravitational interaction). The original theory of GR proposed by Einstein does not admit a nonzero mass for the graviton and so to incorporate this into the Einsten's gravity it is necessary to deform it. As far as we know, a nonzero mass for the graviton may be implemented by adding a nondynamical background metric in GR, as proposed by Visser \citep{Visserothers} or within the framework of $f\left(R\right)$ gravity \citep{StabileCapozziello}. In these types of massive gravity theories, i. e. bimetric theory of gravitation and extended theories of gravity, Yukawa-like corrections to the Newtonian gravitation potential arise naturally within the context of these theories and are not put in by hand. However, it is worthwhile to emphasize that, from a phenomenological viewpoint and inspired in supergravity theories \citep{Hut(1981)}, Yukawa-like terms were introduced long ago by deforming the gravitational Newtonian potential in order to explain the flat rotation curves of spiral galaxies, without assuming the existence of dark matter \citep{Sanders(1984)}.

The purpose of this work is twofold. Firstly, we would like to solve a technical problem and some inconsistencies derived from it, observed
in \citep{Benisty(2022)}, and secondly, we aim to construct a Keplerian-type parametrization for the solution of the two-body problem in Buckingham gravity. In order to contextualize the problem and understand it, it is vital to recall Bertrand's theorem \citep{Bertrand(1873)}. This asserts that the only central forces that would lead to closed paths are those that obey the $1/r^{2}-$force law or the $r-$force law. In terms of the potential energy, this theorem can be rephrased saying that closed paths are only possible when the potential energy
varies as $\sim1/r$ or as $\sim r^{2}$. 

The Yukawa-type gravitational potential studied in \citep{Benisty(2022)} has the form
\begin{eqnarray}
v\left(r\right) & = & -\frac{\kappa}{r}\left(1+\alpha\text{e}^{-mr}\right)\nonumber \\
 & = & -\frac{\kappa\left(1+\alpha\right)}{r}+\kappa\alpha m-\frac{\kappa\alpha m^{2}}{2}r+\cdots,\label{eq:I.1}
\end{eqnarray}
where $\kappa\equiv GM$, with $G$ standing for the gravitational Newtonian constant and $M=m_{1}+m_{2}$ for the total mass of the binary. The dimensionless constant $\alpha$ is the Yukawa strength and $m$ is the Yukawa mass. The second line in (\ref{eq:I.1}) shows the first three terms of the series expansion of the Yukawa contribution, i.e., up to $m^{2}r^{2}$ order, along with the gravitational Newtonian potential. Notice that if we turn off the Yukawa mass, $m\rightarrow0$, in (\ref{eq:I.1}), the residual Yukawa contribution, i.e., the coupling constant $\alpha$, modifies only the gravitational constant $G$, $G\rightarrow G\left(1+\alpha\right)$, but not the $1/r$ potential law. This means that, according to Bertrand's theorem, the Yukawa strength $\alpha$ alone could not affect the position of the pericenter of the ``unperturbed'' closed path. Similarly, the second $r$ constant term, $\kappa\alpha m$, also could not affect the pericenter position, for it may be totally absorbed by redefining the own potential $v\left(r\right)$. However, the third contribution in (\ref{eq:I.1}) really modifies the $1/r $ potential law and, therefore, according to Bertrand's theorem, there  is  an expected displacement of the pericenter due to it, and consequently the path of the finite motion will no longer be closed.

This simple analysis shows that the formula for the advance of the pericenter, up to this order of approximation, should have the structure 
$\delta\theta_{Yuk}=\zeta\left(e\right)\alpha m^{2}a^{2}$, where $\zeta\left(e\right)$ is a function of the eccentricity $e$ and $a$ is the major semiaxis (note that the presence of $a^{2}$ in $\delta\theta_{Yuk}$ comes from the dimensionless requirement of it). This result diverges from that found in \citep{Benisty(2022)}, where the author derives an expression for the pericenter advance (see Eq. (15) in \citep{Benisty(2022)}) that comprises two pieces, one that depends only on $\alpha$ and so violates Bertrand's theorem, as argued above, and the other that is dependent on $f\equiv\alpha m^{2}a^{2}$. In this work, we solve this conflict (and some other inconsistencies in the approximations used in \citep{Benisty(2022)}) by revisiting the Keplerian-type parametrization of the two-body problem in Yukawa
gravity. 

This paper is organized as follows. In Sec.\ref{sec:Sec2}, we give an overview of the Kepler problem. In Sec.\ref{sec:Sec3}, the Keplerian-type
parametrization of the two-body problem in Yukawa gravity is revisited. Here we point out the problems with the approximation used in \citep{Benisty(2022)} and solve them. In particular, the correct formula for the pericenter advance [see Eq. \ref{eq:III.15}] that supports our claims is derived as a byproduct of this parametrization. In Sec. \ref{sec:Sec4}, the Buckingham potential \citep{Buckingham(1933)} is coupled with the gravitational Newtonian potential in order to construct what we shall call Buckingham gravity. The two-body problem is investigated in this kind of gravity by finding a Keplerian-type parametrization for the solution of it. In Sec. \ref{sec:Sec5}, the models for the advance of the pericenter both in Yukawa gravity and in Buckingham gravity are compared with Solar System and S2 star data for determining the constraints on the free parameters. Section \ref{sec:Sec6} contains our main results. In Appendix \ref{sec:AppendixA}, our results for the advance of the pericenter both in Yukawa gravity and in Buckingham gravity are confirmed by using the Landau and Lifshitz's method. Finally, Appendix \ref{sec:AppendixB} contains some functions defined in Buckingham gravity.

\section{An overview of the Kepler problem \protect\label{sec:Sec2}}

As is well known, the motion of two bodies about their center of mass (CM), with interaction potential energy, $U$, that depends only on
the distance between the bodies, i. e. $U=U\left(r\right)$, is equivalent to the motion of one body in this potential energy. This equivalent
one-body problem is straightforwardly solved by invoking the laws of conservation of energy and angular momentum \footnote{These conservation laws can easily found from the Lagrangian formalism of the problem: $\mathcal{L}=\frac{\mu}{2}\left(\dot{r}^{2}+r^{2}\dot{\theta}^{2}\right)-U\left(r\right)$ The angular momentum conservation follows immediately from the fact that the angular coordinate $\theta$ is cyclic, i.e., it does not appear explicitly in the Lagrangian, and therefore its generalized momentum $p_{\theta}=\partial\mathcal{L}/\partial\dot{\theta}$ should be conserved: $p_{\theta}=\mu r^{2}\dot{\theta}=L$. The constancy of the energy, on the other hand, can be established by showing explicitly, with the help of the radial equation of motion $\mu\ddot{r}-\mu r\dot{\theta}^{2}+dU/dr=0$ and the angular momentum conservation, that $dE/dt$ is indeed equal to zero.},
\begin{equation}
E\equiv\mu\epsilon=\frac{\mu}{2}\left(\dot{r}^{2}+r^{2}\dot{\theta}^{2}\right)+\mu v\left(r\right),\qquad\qquad\qquad L\equiv\mu l=\mu r^{2}\dot{\theta},\label{eq:II.1}
\end{equation}
where $r$ and $\theta$ stand for the polar coordinates in the plane of motion and $\mu$ is the reduced mass of the binary system, $\mu=m_{1}m_{2}/M$, with $M=m_{1}+m_{2}$ as the total mass. Notice that $\epsilon$, $l$ and $v(r)$ are, respectively, the energy, angular momentum and potential
energy per unit reduced mass.

Solving the latter equation in (\ref{eq:II.1}) for the angular velocity $\dot{\theta}$, $\dot{\theta}=l/r^{2}$, and substituting it in the
energy equation, we obtain 
\begin{equation}
\epsilon=\frac{1}{2}\dot{r}^{2}+\frac{l^{2}}{2r^{2}}+v\left(r\right).\label{eq:II.2}
\end{equation}
This equation defines the radial motion of the ``equivalent'' body within an effective potential: $v_{eff}(r)=v\left(r\right)+l^{2}/(2r^{2})$,
where the quantity $l^{2}/(2r^{2})$ is called the centrifugal potential. Note that setting $\dot{r}=0$ in (\ref{eq:II.2}) and solving it
for $r$, i.e., the equation $v_{eff}\left(r\right)=\epsilon$, one finds the \textit{turning points} of the motion. If this equation
has only one root $r=r_{min}$, with $r\geq r_{min}$, the motion is infinite, while if there are two roots, $r_{min}$ and $r_{max}$,
with $r_{min}\leq r\leq r_{max}$, the motion is finite. In this case, the path of the body lies completely in the annulus limited by the
circles $r=r_{min}$ and $r=r_{max}$. The equation of the path is obtained by solving (\ref{eq:II.2}) for $\dot{r}$ and dividing by
$\dot{\theta}=l/r^{2}$: $\frac{dr}{d\theta}=\frac{r^{2}}{l}\sqrt{2\left(\epsilon-v\left(r\right)\right)-l^{2}/r^{2}}$. From this result, one sees that if the body goes from $r_{min}$ to $r_{max}$ and then back to $r_{min}$, the polar angle $\theta$ varies by
\begin{equation}
\Delta\theta=2\int_{r_{min}}^{r_{max}}\frac{ldr}{r^{2}\sqrt{2\left(\epsilon-v\left(r\right)\right)-l^{2}/r^{2}}}.\label{eq:II.3}
\end{equation}

It is worth mentioning that the path in a finite motion is not necessarily closed. Indeed, from (\ref{eq:II.3}) we can see that the path of
the body will be closed if and only if $\Delta\theta=2\pi m/n$, where $m$ and $n$ are integers. According to Bertrand's theorem \citep{Bertrand(1873)}, the only central fields that result in closed paths for all finite motions are those in which the potential energy is inversely proportional to $r$ $\left(v\sim1/r\right)$, or is proportional to $r^{2}$ $\left(v\sim r^{2}\right)$. The first case corresponds to a force inversely proportional to $r^{2}$, say, the Newtonian gravitational attraction or the Coulomb electrostatic interaction, while the other case corresponds to a force proportional to $r$, the Hooke's law (the space oscillator).

From now on, we focus our attention on the Newtonian gravitational field, where $U\left(r\right)=-Gm_{1}m_{2}/r$ and so $v\left(r\right)=U\left(r\right)/\mu=-\kappa/r$, with $\kappa\equiv GM$. Substituting $v\left(r\right)$ in the equation found for $dr/d\theta$ and integrating it, we get the orbit equation, a conic section with one focus at the origin,
\begin{equation}
\frac{p}{r}=1+e\cos\theta,\label{eq:II.4}
\end{equation}
where $p=l^{2}/\kappa$ is the semilatus rectum of the orbit and $e=\sqrt{1+2\epsilon l^{2}/\kappa^{2}}$ is its eccentricity. For $\epsilon<0$,
the motion is finite and closed, with the orbit being an ellipse $\left(e<1\right)$, with the major and minor semiaxes given respectively by $a=p/(1-e^{2})=\kappa/\left(2\left|\epsilon\right|\right)$ and $b=p/\sqrt{1-e^{2}}=l/\sqrt{2\left|\epsilon\right|}$. The apsides
of the elliptical orbit, according to (\ref{eq:II.4}), i. e., the pericenter (periastron) and apocenter (apastron) points occur at $\theta=0$
and $\theta=\pi$, respectively, with apsidal distances given by
\begin{equation}
r_{min}=a\left(1-e\right)\,\,\,\,\,\,\,\,\,\,\,\,\,\,\,\,\,\,\,\,\,\,\,\,\,\,\,\,\,\,r_{max}=a\left(1+e\right).\label{eq:II.5}
\end{equation}
It should be noted that the orbit turns out to be a circle $\left(e=0\right)$ when $\epsilon$ equals the minimum value of the effective potential
$v_{eff}=-\kappa/r+l^{2}/\left(2r^{2}\right)$, i. e., $\epsilon=v_{eff,min}=-\kappa^{2}/\left(2l^{2}\right)$. For $\epsilon\geq0$, the motion is infinite and the orbit becomes a parabola $\left(e=1\right)$ for $\epsilon=0$ and a hyperbola $\left(e>1\right)$ for $\epsilon>0$. Since we are interested in the motion of two nonspinning compact objects (i.e., binary star systems) moving in an ``eccentric'' orbit, we restrict our study to the finite motion case $\left(v_{eff,min}<\epsilon<0\right)$.

To determine the radial distance $r$ as function of time, we first solve (\ref{eq:II.2}) for $\dot{r}$, $\dot{r}\equiv dr/dt=\sqrt{2\left(\epsilon+\kappa/r\right)-l^{2}/r^{2}}$, where we have set $v\left(r\right)=-\kappa/r$, and then integrate it by changing the integration variable $r$ to the auxiliary one $\eta$ (called the eccentric anomaly) which is defined by the relation
\begin{equation}
r=a\left(1-e\cos\eta\right).\label{eq:II.6}
\end{equation}
The result of doing this is the well-known Kepler's equation,
\begin{equation}
\frac{2\pi}{T}\left(t-t_{0}\right)=\eta-e\sin\eta,\label{eq:II.7}
\end{equation}
where $t_{0}$ is termed the time of pericenter passage and $T=2\pi\sqrt{a^{3}/\kappa}$ is the orbital period. The quantity that appears on the left-hand side of (\ref{eq:II.7}), i.e., $\mathcal{M}\equiv\frac{2\pi}{T}\left(t-t_{0}\right)$, is referred to as the mean anomaly. On the other hand, to find the polar angle $\theta$ (referred to in this context as the true anomaly) in terms of the eccentric anomaly $\eta$, one plugs $r$ given by (\ref{eq:II.6}) into the orbit equation (\ref{eq:II.4}) and simplifies it to give $\cos\theta=\left(\cos\eta-e\right)/\left(1-e\cos\eta\right)$. An alternate form of writing this expression can be obtained by using the trigonometric identity $\tan^{2}\theta/2=\left(1-\cos\theta\right)/\left(1+\cos\theta\right)$. The result is
\begin{equation}
\theta=2\text{tan}^{-1}\left[\sqrt{\frac{1+e}{1-e}}\tan\frac{\eta}{2}\right].\label{eq:II.8}
\end{equation}

Here some remarks are in order. First, the parametrization of the radial and angular coordinates, $r$ and $\theta$, respectively,
in terms of the eccentric anomaly $\eta$, given by the equations (\ref{eq:II.6}) and (\ref{eq:II.8}), is known in the literature
as the Keplerian parametrization for Newtonian accurate orbital motion of a binary system \citep{Memmesheimeretal(2004)}. It should be stressed that in the Newtonian approximation to gravity (where the inverse-square law is valid), there is no advance of the pericenter of the elliptical orbit. This is guaranteed by Bertrand's theorem, a keystone in celestial mechanics \citep{Bertrand(1873)}.
Another way to see this is by showing explicitly that the so-called Laplace-Runge-Lenz vector, $\overrightarrow{\mathcal{L}}=\dot{\overrightarrow{r}}\times\overrightarrow{l}-\kappa\,\overrightarrow{r}/r$, is a constant vector (for an inverse square central force, independent of the value of $\kappa$), of magnitude $\left|\overrightarrow{\mathcal{L}}\right|=\kappa e$, and is directed from the focus toward the pericenter point. In this way the position of the pericenter does not change and so the orbit closes. Second, in the post-Newtonian (PN) approximations to gravity \citep{Weinberg(1972)}, where the equations of motion of the binary system are described as corrections to the Newtonian ones in terms of the dimensionless parameter, $v^{2}/c^{2}\sim GM/c^{2}r$, the Keplerian parametrization is extended by compelling the radial parametrization (\ref{eq:II.6}) to maintain its structure, adjusting adequately the parameters $a$ and $e$ {(}see \citep{Memmesheimeretal(2004)}
and references therein{)}.

\section{Yukawa gravity (a revisited version) \protect\label{sec:Sec3}}

In this section, the Keplerian-type parametrization constructed in \citep{Benisty(2022)} for the solution of the two-body problem
in Yukawa gravity is revisited. This is done for two reasons. Firstly, to review the parametrization techniques (that will be used in the
next section) used in \citep{DamourDeruelle(1985)} for finding an analytic parametric solution of the two-body problem in the 1PN approximation to gravity and recently implemented in Yukawa gravity \citep{Benisty(2022)}, and secondly, to sort out some inconsistencies observed, in particular, in the pericenter advance formulae, which evidently violate Bertrand's theorem.

Our starting point is the energy equation [\ref{eq:II.2}]. Setting $v\left(r\right)$ as
\begin{equation}
v\left(r\right)=-\frac{\kappa}{r}\left(1+\alpha\text{e}^{-mr}\right)-\frac{\kappa l^{2}}{c^{2}r^{3}},\label{eq:III.1}
\end{equation}
we get
\begin{equation}
\epsilon=\frac{1}{2}\dot{r}^{2}-\frac{\kappa}{r}\left(1+\alpha\text{e}^{-mr}\right)+\frac{l^{2}}{2r^{2}}\left(1-\frac{2\kappa}{c^{2}r}\right).\label{eq:III.2}
\end{equation}
Note that in (\ref{eq:III.1}), and thus in (\ref{eq:III.2}), we are adding the general relativistic correction, $v_{GR}=-\kappa l^{2}/\left(c^{2}r^{3}\right)$, to the gravitational Newtonian potential, which corresponds to the 1PN approximation to gravity \citep{Weinberg(1972),Zee(2013)}.
Since the leading term in the expansion of the Yukawa potential that might give rise to a pericenter advance is of second order in $m$,
we shall truncate the exponential expansion, considering $mr\ll1$, at this order, i.e., $\text{e}^{-mr}\approx1-mr+m^{2}r^{2}/2$.

By imposing the boundary conditions $\dot{r}[r_{min,max}]=0$, with $r_{min}$ and $r_{max}$ given in (\ref{eq:II.5}), in the energy Eq. (\ref{eq:III.2}), one finds the energy $\epsilon$ and the angular momentum $l$ in terms of the $a$ and $e$ orbital parameters
\begin{equation}
\epsilon=-\frac{\kappa}{2a}\left\{ 1+\alpha-2\alpha ma+\frac{f}{2}\left(3+e^{2}\right)-\beta\left(1-e^{2}\right)\left[1+\alpha-\frac{f}{2}\left(1-e^{2}\right)\right]\right\} \label{eq:III.3}
\end{equation}
and 
\begin{equation}
\frac{l^{2}}{\kappa a\left(1-e^{2}\right)}=\left[1+\beta\left(3+e^{2}\right)\right]\left[1+\alpha-\frac{f}{2}\left(1-e^{2}\right)\right],\label{eq:III.4}
\end{equation}
where $f\equiv\alpha m^{2}a^{2}$ and $\beta\equiv\kappa/\left[ac^{2}\left(1-e^{2}\right)\right]$. The use of the boundary conditions, $\dot{r}[r_{min,max}]=0$, implies that we are assuming that the motion is finite (limited to the annulus defined by the circles $r=r_{min,max}$ ), but no longer closed. To be consistent with our approximation, terms up to second order in $ma\ll1$ are retained both in (\ref{eq:III.3}) and in (\ref{eq:III.4}). It is important to stress that this is not done in \citep{Benisty(2022)}, where the $\epsilon$ equation is truncated up to first order in $ma$, while the $l$ equation goes up to second order in $ma$.

To construct a solution of the Kepler type, one introduces the eccentric anomaly $\eta$ by means of the radial Eq. (\ref{eq:II.6}),
noting that the orbital parameters $a$ and $e$ are defined implicitly by (\ref{eq:III.3}) and (\ref{eq:III.4}). Substituting (\ref{eq:II.6})
into (\ref{eq:III.2}) and eliminating $\epsilon$ and $l$, with the help of the relations (\ref{eq:III.3}) and (\ref{eq:III.4}), one finds the differential equation for $\eta$
\begin{equation}
\frac{\dot{\eta}^{-1}-\dot{\eta}^{-1}\left(\beta=0\right)}{\sqrt{a^{3}/\kappa}}=\frac{\beta}{8}\left(1-e^{2}\right)\left(e\text{cos}\eta-3\right)\left[2\left(\alpha-2\right)+f\left(e^{2}+6e\text{cos}\eta-7\right)\right],\label{eq:III.5}
\end{equation}
where 
\begin{equation}
\dot{\eta}^{-1}\left(\beta=0\right)=\sqrt{\frac{a^{3}}{\kappa}}\frac{\left(1-e\text{cos}\eta\right)}{4}\left[2\left(2-\alpha\right)+f\left(3-e^{2}\right)-2ef\text{cos}\eta\right].\label{eq:III.6}
\end{equation}
The solution of (\ref{eq:III.5}), after some algebraic manipulations, can be put in the form
\begin{equation}
\frac{2\pi}{T}\left(t-t_{0}\right)=\eta-e_{t}\text{sin}\eta+\frac{e^{2}f}{8}\text{sin}\left(2\eta\right)\label{eq:III.7}
\end{equation}
where the period $T$ and the time eccentricity $e_{t}$ are given by
\begin{equation}
T=2\pi\sqrt{\frac{a^{3}}{\kappa}}\left[1-\frac{\alpha}{2}+\frac{3f}{4}+\frac{3\beta}{8}\left(1-e^{2}\right)\left(4+7f-2\alpha\right)\right]\label{eq:III.8}
\end{equation}
and 
\begin{equation}
\frac{e_{t}}{e}=1+\frac{\left(2-e^{2}\right)f}{4}-\beta\left(1-e^{2}\right)\left[1-\frac{\left(2+e^{2}\right)f}{4}\right].\label{eq:III.9}
\end{equation}
Equation (\ref{eq:III.7}) is the extension of the Kepler one [see Eq. (\ref{eq:II.7})] for the two-body problem in Yukawa gravity.
Once the structure of the original Kepler equation is broken by the presence of the $f$ Yukawa term in (\ref{eq:III.7}), this kind of
parametrization will be called ``generalized quasi-Keplerian'' parametrization, following the nomenclature adopted in \citep{DamourSchafer(1988)}. 

Finally, the $\eta$ parametrization of the true anomaly $\theta$ is obtained by means of the relation $d\theta/d\eta=\dot{\theta}/\dot{\eta}=l/\left(r^{2}\dot{\eta}\right)$. Substituting $r$ and $\dot{\eta}^{-1}$ given in (\ref{eq:II.6}) and (\ref{eq:III.5}), respectively, and keeping only the relevant terms, i.e., second order in $ma\ll1$ and first order in $\beta\ll1$, we get
{\footnotesize
\begin{equation}
\frac{d\theta}{d\eta}-\frac{d\theta}{d\eta}\left(\beta=0\right)=\frac{\beta\sqrt{1-e^{2}}}{\left(1-e\rho\right)^{2}}\left\{ 3-e^{2}+\left(3-2e^{2}\right)f-\frac{e}{2}\left[4+\left(9-5e^{^{2}}\right)f\right]\rho+\frac{e^{2}\left(3-e^{2}\right)f}{2}\rho^{2}\right\} \label{eq:III.10}
\end{equation}
} where
\begin{equation}
\frac{d\theta}{d\eta}\left(\beta=0\right)=\frac{\sqrt{1-e^{2}}}{1-e\rho}\left(1+\frac{f}{2}-\frac{ef}{2}\rho\right),\label{eq:III.11}
\end{equation}
with $\rho=\text{cos}\eta$. 

The solution of (\ref{eq:III.10}) can be cast in the form
\begin{equation}
\frac{2\pi\left(\theta-\theta_{0}\right)}{\Phi}=\nu+\frac{\sqrt{1-e^{2}}}{2}\left[1+\left(3-e^{2}\right)\beta\right]f\left(\eta-\nu\right)+\frac{e\beta\sqrt{1-e^{2}}\,\text{sin}\eta}{\left(1-e\text{cos}\eta\right)}\label{eq:III.12}
\end{equation}
where
\begin{equation}
\frac{\Phi}{2\pi}=1+\frac{\sqrt{1-e^{2}}}{2}\left[1+\left(3-e^{2}\right)\beta\right]f+\frac{3}{2}\left[2+\left(1-e^{2}\right)f\right]\beta,\label{eq:III.13}
\end{equation}
with $\nu=2\text{tan}^{^{-1}}\left[\sqrt{\frac{1+e}{1-e}}\text{tan}\frac{\eta}{2}\right]$. These results, in particular, Eqs. (\ref{eq:III.10}), (\ref{eq:III.12}) and (\ref{eq:III.13}), differ from those obtained in \citep{Benisty(2022)} (see, for comparison,  Eqs. (20), (22), (23) and (24) in \citep{Benisty(2022)}). It should be noted that the Yukawa strength $\alpha$ appears in these equations only through the definition of $f,$ i. e., in the combination $f=\alpha m^{2}a^{2}$, as expected.

\begin{figure}[H]
\begin{centering}
\includegraphics{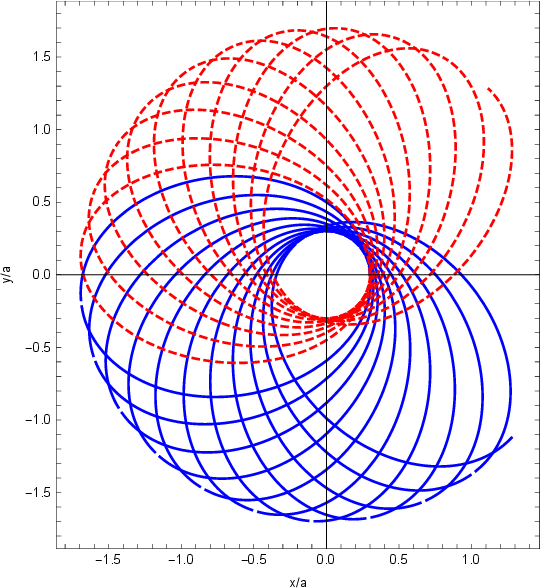}
\par\end{centering}
\caption{\protect\label{fig:1} Paths of the binary for $\alpha>0$ (the solid path) and for $\alpha<0$ (the dashed path) in Yukawa gravity. The
Yukawa potential produces a precession of the pericenter given by (\ref{eq:III.15}).}
\end{figure}

From Eq. (\ref{eq:III.13}), the pericenter precession rate reads
\begin{equation}
\delta\theta_{Yuk-GR}=\frac{\sqrt{1-e^{2}}}{2}\left[1+\left(3-e^{2}\right)\beta\right]f+\frac{3}{2}\left[2+\left(1-e^{2}\right)f\right]\beta.\label{eq:III.14}
\end{equation}
This is the right expression for the advance of the pericenter. This depends only on $\alpha$ by means of the combination $f=\alpha m^{2}a^{2}$ and, therefore, does not violate Bertrand's theorem. 
If one switches off the relativistic correction in (\ref{eq:III.14}), taking $\beta\rightarrow0$, one gets the pure Yukawa contribution for the advance of the pericenter
\begin{equation}
\delta\theta_{Yuk}=\frac{\sqrt{1-e^{2}}}{2}f.\label{eq:III.15}
\end{equation}
This result respects Bertrand's theorem and is completely different from that found in \citep{Benisty(2022)} (see Eq. (15) in \citep{Benisty(2022)}).
On the other hand, if one switches off the Yukawa contribution in (\ref{eq:III.14}), 
taking $f\rightarrow0$, one recovers the well-known relativistic result for the advance of the pericenter, i. e., $\delta\theta_{GR}=3\beta=3GM/\left[ac^{2}\left(1-e^{2}\right)\right]$. This does not occur in the result of (24) found in \citep{Benisty(2022)}.
The paths of the binary for $\alpha>0$ (the solid path) and for $\alpha<0$ (the dashed path) in Yukawa gravity $\left(f\neq0,\,\beta=0\right)$
are shown in Fig. \ref{fig:1}. These paths are obtained from the relations $x=r\cos\theta$ and $y=r\sin\theta$ by eliminating $r$
and $\theta$ in terms of $\eta$ with the help of (\ref{eq:II.6}) and (\ref{eq:III.12}), respectively. 

The results for the advance of the pericenter are corroborated in Appendix \ref{sec:AppendixA} by using the Landau and Lifshitz's method \citep{LandauLifshitz(1976)}.

\section{Buckingham gravity \protect\label{sec:Sec4}}

\begin{figure}[H]
\begin{centering}
\includegraphics{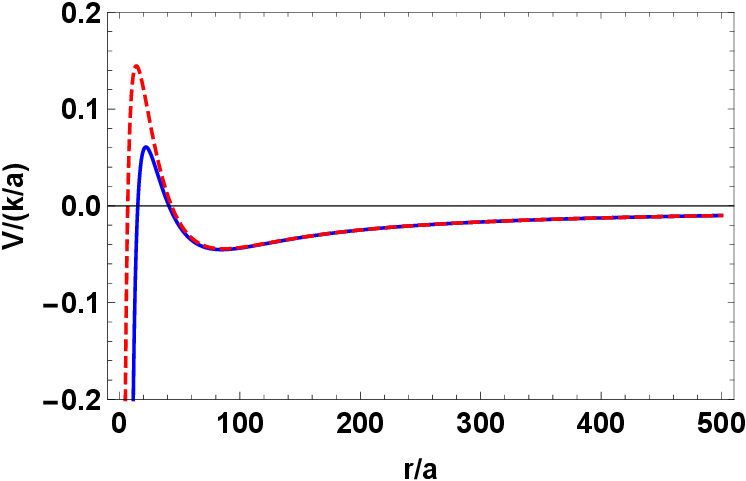}
\par\end{centering}
\caption{\protect\label{fig:2} The dashed line represents the Buckingham-Newtonian potential with the centrifugal potential, while the solid line represents the total potential, including the relativistic correction.}
\end{figure}

This section is devoted to finding a Keplerian-type parametrization of the two-body problem in what we call Buckingham gravity. The name
derives from the fact that we couple the Buckingham potential \citep{Buckingham(1933)}, $v_{Buck}=A\exp\left(-Br\right)-C/r^{6}$, to the Newtonian potential of gravity. Our starting point is once again the energy Eq. (\ref{eq:II.2}), with
\begin{equation}
v\left(r\right)=-\frac{\kappa}{r}\left(1-\alpha r\text{e}^{-mr}+\frac{\gamma}{r^{5}}\right)-\frac{\kappa l^{2}}{c^{2}r^{3}},\label{eq:IV.1}
\end{equation}
that is,
\begin{equation}
\epsilon=\frac{1}{2}\dot{r}^{2}-\frac{\kappa}{r}\left(1-\alpha r\text{e}^{-mr}+\frac{\gamma}{r^{5}}\right)+\frac{l^{2}}{2r^{2}}\left(1-\frac{2\kappa}{c^{2}r}\right).\label{eq:IV.2}
\end{equation}
Note that the constants $\alpha$, $m$, and $\gamma$ are not dimensionless: 
$\left[\alpha\right]=\left[m\right]=L^{-1}$ and $\left[\gamma\right]=L^{5}$.
The last term in (\ref{eq:IV.1}), like in Yukawa gravity, corresponds to the 1PN correction to Newtonian gravity. The Buckingham-Newtonian
potentials (in dimensionless units) with and without the relativistic correction are shown in Fig. \ref{fig:2}.

In what follows, we pursue the same approach as in the preceding section. Thus, we first find the energy $\epsilon$ and angular momentum $l$
(per unit reduced mass) in terms of the $a$ and $e$ parameters by applying the boundary conditions $\dot{r}\left[r_{min,max}\right]=0$
in (\ref{eq:IV.2}). The relations found for $\epsilon$ and $l$ are
\begin{eqnarray}
\epsilon & = & -\frac{\kappa}{2a}\left\{ 1-2\alpha a+\tilde{f}\left(3+e^{2}\right)-\frac{4\left(1+e^{2}\right)}{\left(1-e^{2}\right)^{4}}\frac{\gamma}{a^{5}}-\beta\left(1-e^{2}\right)\left[1-\tilde{f}\left(1-e^{2}\right)+\right.\right.\nonumber \\
 &  & \left.\left.+\frac{2\left(3+10e^{2}+3e^{4}\right)}{\left(1-e^{2}\right)^{5}}\frac{\gamma}{a^{5}}\right]\right\} \label{eq:IV.3}
\end{eqnarray}
and
\begin{equation}
\frac{l^{2}}{\kappa a\left(1-e^{2}\right)}=\left[1+\beta\left(3+e^{2}\right)\right]\left[1-\tilde{f}\left(1-e^{2}\right)+\frac{2\left(3+e^{2}\right)\left(1+3e^{2}\right)}{\left(1-e^{2}\right)^{5}}\frac{\gamma}{a^{5}}\right]\label{eq:IV.4}
\end{equation}
where $\tilde{f}\equiv\alpha ma^{2}$ and $\beta=\kappa/\left[ac^{2}\left(1-e^{2}\right)\right]$ are dimensionless constants. It is worth emphasizing that in the expansion of the exponential in (\ref{eq:IV.2}) we are keeping terms only up to first order in $mr\ll1$, that is, 
$\text{e}^{-mr}\approx1-mr$. Since there is an $r \text{factor}$ multiplying the exponential in the Buckingham contribution to gravity [see Eq.(\ref{eq:IV.1})], it seems that this approximation is sufficient to see a possible change in the position of the pericenter. 

The next step is to introduce the eccentric anomaly $\eta$ by means of the defining Eq. (\ref{eq:II.6}). Plunging this into (\ref{eq:IV.2})
and eliminating $\epsilon$ and $l$ with the aid of (\ref{eq:IV.3}) and (\ref{eq:IV.4}), one gets
\begin{equation}
\frac{\dot{\eta}^{-1}-\dot{\eta}^{-1}\left(\beta=0\right)}{\sqrt{a^{3}/\kappa}\left(1-e^{2}\right)}=\frac{\beta}{4}\left(3-e\rho\right)\left\{ 2+\tilde{f}\left(7-e^{2}-6e\rho\right)+\frac{\left[2F\left(e,\rho\right)-3\tilde{f}G\left(e,\rho\right)\right]}{\left(1-e^{2}\right)^{5}\left(1-e\rho\right)^{4}}\frac{\gamma}{a^{5}}\right\} \label{eq:IV.5}
\end{equation}
where
\begin{equation}
\frac{\dot{\eta}^{-1}\left(\beta=0\right)}{\sqrt{a^{3}/\kappa}\left(1-e\rho\right)}=1+\frac{\tilde{f}}{2}x+\frac{\left(2+3\tilde{f}x\right)\left[4\left(1-e\rho\right)-e^{2}\rho'{}^{2}\right]\left[3+e^{4}-2e\left(1+e^{2}\right)\left(2-e\rho\right)\rho\right]}{2\left(1-e^{2}\right)^{4}\left(1-e\rho\right)^{4}}\frac{\gamma}{a^{5}}\label{eq:IV.6}
\end{equation}
{\small with $x\equiv3-e^{2}-2e\rho$ and $\rho'=d\rho/d\eta=-\text{sin}\eta$.
The functions $F\left(e,\rho\right)$ and $G\left(e,\rho\right)$ in (\ref{eq:IV.5}) are given in Appendix \ref{sec:AppendixB}, see
Eqs. (\ref{eq:AppB.1}) and (\ref{eq:AppB.2}), respectively. }{\small\par}

The solution of (\ref{eq:IV.5}) can be cast in the form
\begin{eqnarray}
\frac{2\pi\left(t-t_{0}\right)}{T} & = & \eta-\tilde{e}_{t}\text{sin}\eta+\frac{\tilde{f}e^{2}}{4}\text{sin}2\eta-\frac{3\left[4+e^{2}+5\left(4+3e^{2}\right)\beta\right]\gamma/a^{5}}{2\left(1-e^{2}\right)^{7/2}}\left(\eta-\nu\right)\nonumber \\
 &  & +\frac{1}{8\left(1-e^{2}\right)^{4}}\frac{\gamma/a^{5}}{\left(1-e\rho\right)^{3}}\sum_{n=1}^{4}e^{n}F_{n}\left(e,\beta\right)\text{sin}n\eta\label{eq:IV.7}
\end{eqnarray}
where the period $T$ and the time eccentricity $\tilde{e}_{t}$ are defined by
\begin{eqnarray}
T & = & 2\pi\sqrt{\frac{a^{3}}{\kappa}}\left\{ 1+\frac{3}{2}\tilde{f}+\frac{\beta}{4}\left(1-e^{2}\right)\left(6+21\tilde{f}\right)+\frac{6\gamma/a^{5}}{\left(1-e^{2}\right)^{9/2}}\left[1+\sqrt{1-e^{2}}\left(1+e^{2}\right)\right.\right.\nonumber \\
 &  & \left.\left.-\frac{1}{4}\left(3+e^{2}\right)e^{2}+\frac{\beta}{8}\left[52+e^{2}\left(3e^{4}-24e^{2}-31\right)+8\left(4-e^{4}+5e^{2}\right)\sqrt{1-e^{2}}\right]\right]\right\} \label{eq:IV.8}
\end{eqnarray}
\begin{equation}
\frac{\tilde{e}_{t}}{e}=1+\frac{\left(2-e^{2}\right)}{2}\tilde{f}-\beta\left(1-e^{2}\right)\left[1-\frac{\left(2+e^{2}\right)}{2}\tilde{f}\right].\label{eq:IV.9}
\end{equation}
The functions $F_{n}\left(e,\beta\right)$ that appear in (\ref{eq:IV.7}) are given in Appendix \ref{sec:AppendixB}, see Eqs. (\ref{eq:AppB.3})-(\ref{eq:AppB.6}). {\small For simplicity, $\alpha m\text{\ensuremath{\gamma}}-$terms have been dropped in (\ref{eq:IV.7}).}{\small\par}

The true anomaly $\theta$ in terms of the eccentric anomaly $\eta$ is found with the help of the relation $d\theta/d\eta=l/\left(r^{2}\dot{\eta}\right)$. Thus, after substituting (\ref{eq:II.6}) and (\ref{eq:IV.5}) in it, and performing the necessary algebraic manipulations, one obtains
{\small
\begin{equation}
\frac{d\theta}{d\eta}-\frac{d\theta}{d\eta}\left(\beta=0\right)=\frac{\beta\sqrt{1-e^{2}}}{\left(1-e\rho\right)^{2}}\left\{ x+\tilde{f}\left(1-e\rho\right)\tilde{x}+\frac{\tilde{x}\,\gamma/a^{5}}{8\left(1-e^{2}\right)^{5}\left(1-e\rho\right)^{4}}\sum_{n=0}^{4}e^{n}P_{n}\left(e\right)\text{cos}n\eta\right\} \label{eq:IV.10}
\end{equation}
where $x\equiv3-e^{2}-2e\rho,$} $\tilde{x}=6-4e^{2}-\left(3-e^{2}\right)e\rho$,
\begin{equation}
\frac{d\theta}{d\eta}\left(\beta=0\right)=\frac{\sqrt{1-e^{2}}}{1-e\rho}\left(1+\tilde{f}\left(1-e\rho\right)\right)+\frac{\gamma/a^{5}}{8\left(1-e^{2}\right)^{9/2}\left(1-e\rho\right)^{5}}\sum_{n=0}^{4}e^{n}P_{n}\left(e\right)\text{cos}n\eta\label{eq:IV.11}
\end{equation}
and the polynomials in $e$, $P_{n}\left(e\right)$, are given in Appendix \ref{sec:AppendixB}, see Eqs. (\ref{eq:AppB.7}) and (\ref{eq:AppB.8}).

The integration of (\ref{eq:IV.10}) can be cast in the form
\begin{eqnarray}
\frac{2\pi\left(\theta-\theta_{0}\right)}{\Phi} & = & \nu+\sqrt{1-e^{2}}\left(1-\beta e^{2}\right)\tilde{f}\left(\eta-\nu\right)+\frac{\beta\sqrt{1-e^{2}}e\text{sin}\eta}{\left(1-e\text{cos}\eta\right)}\left(1-\sqrt{1-e^{2}}\tilde{f}\right)+\nonumber \\
 &  & -\frac{\gamma/a^{5}}{128\left(1-e^{2}\right)^{9/2}\left(1-e\text{cos}\eta\right)^{5}}\sum_{n=1}^{5}e^{n}\tilde{P}_{n}\left(e,\beta\right)\text{sin}n\eta,\label{eq:IV.12}
\end{eqnarray}
where
\begin{eqnarray}
\frac{\Phi}{2\pi} & = & 1+\sqrt{1-e^{2}}\left[1+\beta\left(3-e^{2}\right)\right]\tilde{f}+3\beta\left[1+\left(1-e^{2}\right)\tilde{f}\right]+\frac{15\,\gamma/a^{5}}{8\left(1-e^{2}\right)^{5}}\left[8+12e^{2}\right.\nonumber \\
 &  & \left.+e^{4}+\frac{\beta}{5}\left(240+400e^{2}+12e^{4}-5e^{6}\right)\right],\label{eq:IV.13}
\end{eqnarray}
and the functions $\tilde{P}_{n}\left(e,\beta\right)$ in (\ref{eq:IV.12}) are given in Appendix \ref{sec:AppendixB}, see Eqs. (\ref{eq:AppB.9})-(\ref{eq:AppB.13}).

From this last expression, the pericenter precession rate reads
\begin{eqnarray}
\delta\theta_{Buck-GR} & = & \sqrt{1-e^{2}}\left[1+\beta\left(3-e^{2}\right)\right]\tilde{f}+3\beta\left[1+\left(1-e^{2}\right)\tilde{f}\right]+\frac{15\,\gamma/a^{5}}{8\left(1-e^{2}\right)^{5}}\left[8+12e^{2}\right.\nonumber \\
 &  & \left.+e^{4}+\frac{\beta}{5}\left(240+400e^{2}+12e^{4}-5e^{6}\right)\right].\label{eq:IV.14}
\end{eqnarray}
If we switch off the relativistic contribution in (\ref{eq:IV.14}), by setting $\beta\rightarrow0$, we verify that there is a Buckingham
effect in the advance of the pericenter, namely,
\begin{equation}
\delta\theta_{Buck}=\sqrt{1-e^{2}}\tilde{f}+\frac{15\gamma/a^{5}}{8\left(1-e^{2}\right)^{5}}\left(8+12e^{2}+e^{4}\right).\label{eq:IV.15}
\end{equation}
On the other hand, if the Buckingham contribution is turned off by taking $\tilde{f}\rightarrow0$ and $\gamma\rightarrow0$ in (\ref{eq:IV.14}),
one recovers the relativistic result: $\delta\theta_{GR}=3\beta=3GM/\left[ac^{2}\left(1-e^{2}\right)\right]$.

Like in Yukawa gravity, these results for the advance of the pericenter are confirmed in Appendix \ref{sec:AppendixB} by using the Landau
and Lifshitz's method \citep{LandauLifshitz(1976)}.

\section{Observational tests of modified gravity\protect\label{sec:Sec5}}
Here, we compare the obtained results for Yukawa and Buckingham gravity with observational data, in order to constrain the free parameters of the theories. The data we choose to work here are Solar System data and S2 star data.

The Solar System data we use are from INPOP10a planetary ephemeris from \cite{FiengaEtAl11}, which consists of supplementary advances of perihelia for six planets, from Mercury to Saturn. GRAVITY collaboration has obtained quite precise data from S2, one of many stars orbiting the compact radio source Sgr A$^*$, at the center of the Galaxy, including a measurement of the precession of its orbit \cite{AbuterEtAl20}, which we also use here in order to obtain constraints over our free parameters.

The method we have used to constrain the free parameters with these data, consists of obtaining the $\chi^2$ for each dataset
\begin{align}
    \chi^2_{SS}(\vec{p})&=\sum_{i=1}^6\left[\frac{\delta\theta_{mod,i}(\vec{p})-\delta\theta_{obs,i}}{\sigma_{\delta\theta,i}}\right]^2\\
    \chi^2_{S2}(\vec{p})&=\left[\frac{\delta\theta_{mod}(\vec{p})-\delta\theta_{obs,S2}}{\sigma_{\delta\theta,S2}}\right]^2
\end{align}
where the index $SS$ stands for Solar System, $S2$ for S2 star, $\delta\theta_{mod}$ is the predicted precession from the model, $\delta\theta_{obs}$ is the observed precession and $\vec{p}$ is the vector of free parameters of each model. As $SS$ and $S2$ are statistically independent data, one can obtain combined constraints from simply summing the $\chi^2_i$
\begin{eqnarray}
    \chi^2(\vec{p})=\chi^2_{SS}(\vec{p})+\chi^2_{S2}(\vec{p})
\end{eqnarray}

What are the free parameters $\vec{p}$? In Yukawa gravity, we have two parameters, $\alpha$ and $m$. However, as can be seen from Eq. \eqref{eq:III.14}, these parameters appear only in the definition of the $f$ parameter, i.e., $f=\alpha m^2 a^2$, so, one can not obtain separated constraints for these from perihelia precession data alone. Therefore, here we choose to work with the product
\begin{eqnarray}
 \alpha_Y\equiv\alpha m^2=\frac{f}{a^2}
\end{eqnarray}
as the only ``free'' parameter for Yukawa gravity. In Yukawa gravity, $\alpha$ is dimensionless, while $m$ has an inverse length dimension, $[m]=L^{-1}$. So, $[\alpha_Y]=L^{-2}$.

Something similar happens with Buckingham gravity. In principle, it would have three free parameters, $\alpha$, $m$ and $\gamma$. However, as can be seen from Eq. \eqref{eq:IV.14}, $\alpha$ and $m$ appear only as a product contained in $\tilde{f}\equiv\alpha m a^2$. So, we have to work with the product
\begin{eqnarray}
 \alpha_B\equiv\alpha m=\frac{\tilde{f}}{a^2}
\end{eqnarray}
as a ``free'' parameter. Our vector of free parameters, in this case, is $\vec{p}_B=(\alpha_B,\gamma)$. As mentioned in Sec. \ref{sec:Sec4}, $[\alpha]=[m]=L^{-1}$, in such a way that $[\alpha_B]=L^{-2}$. Recall also that $[\gamma]=L^5$.

It is important to stress that by comparing the models for the advance of the pericenter [see Eqs. \eqref{eq:III.14} and \eqref{eq:IV.14}] with the pericenter precession observations of the Solar System and the S2 star, one can only find out constraints on the products $\alpha_Y\equiv\alpha m^2$ and $\alpha_B\equiv\alpha m$ (along with $\gamma$) in Yukawa gravity and Buckingham gravity, respectively. This claim clearly enters in conflict with the results present in \citep{Benisty(2022)}, where the author found constraints on both $\alpha$ and $m$ Yukawa parameters. Nevertheless, as proved in this work, the model for the advance of the pericenter derived in \citep{Benisty(2022)}, as a byproduct of the Keplerian-like parametrization of the two-body problem in Yukawa gravity, is not correct. This violates Bertrand's theorem: the Yukawa strength $\alpha$ should not appear isolated in the pericenter's advance formula, as occurs in Eqs. (15) and (24) of \citep{Benisty(2022)}.

Following the standard procedure of Bayesian statistics, we assumed flat priors $\pi(\vec{p})=cnst.$ for the free parameters of both models and defined the likelihood as $\mathcal{L}(\vec{p})=Ne^{-\frac{1}{2}\chi^2(\vec{p})}$, where $N$ is a normalization constant. Bayes' theorem states that the posterior distribution of the free parameters is given by $\rho\propto\pi\mathcal{L}$. The posterior encodes all the information about the free parameters given the data. In order to probe the posterior, we have used the python implementation of the affine invariant Markov chain Monte Carlo (MCMC) ensemble sampler, \citep{emcee}. 
In order to analyze and plot the obtained MCMC samples, we have used the package \citep{getdist}. 
Below we show the results for each model.

\subsection{Yukawa gravity}
\begin{figure}[ht]
    \centering
    \includegraphics[width=.6\linewidth]{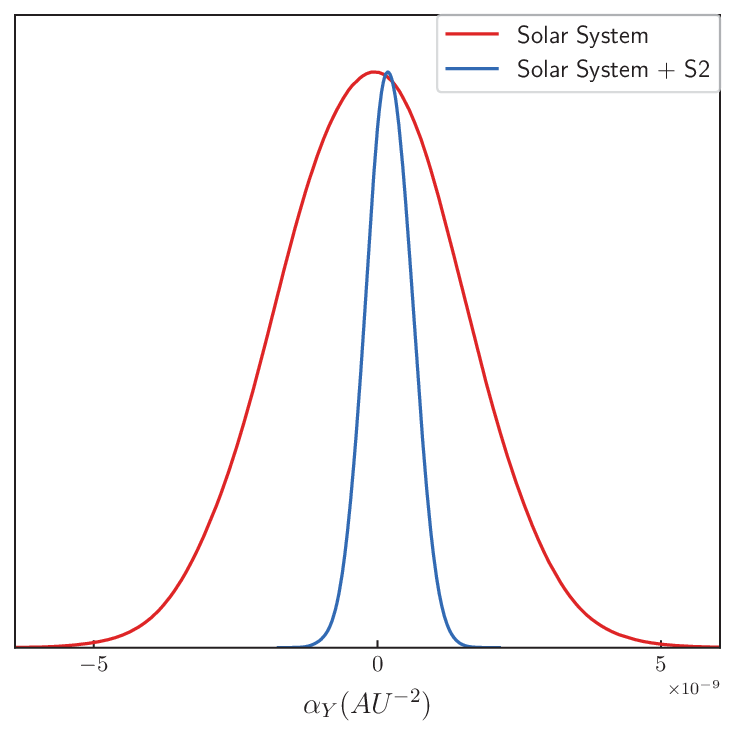}
    \caption{Posterior distributions for the $\alpha_Y$ parameter in Yukawa gravity. In red is the posterior for the Solar System data alone. In blue is the combined posterior from Solar System+S2 data.}
    \label{fig:Yukawa}
\end{figure}

In the case of Yukawa gravity, the constraints over the parameter $\alpha_Y$ can be seen in Fig. \ref{fig:Yukawa}.

As can be seen from this figure, $\alpha_Y$ is quite constrained when  the S2 data are added to the Solar System data. In fact, from Solar System data alone, the best fit for $\alpha_Y$ is negative, while when one adds the S2 constraint, we find $\alpha_Y>0$ at the best fit. Table \ref{tab:Yukawa} below shows the quantitative constraints.

\begin{table}[ht]
    \centering
    \begin{tabular} { l  c c}
 Parameter \hspace{1cm} & \hspace{1cm} SS Constraints \hspace{1cm} & \hspace{1cm} SS+S2 Constraints \hspace{1cm} \\
\hline
{\boldmath$\alpha_Y$} (AU$^{-2}$) & $\left(\,-0.1\pm1.6\pm3.2\,\right)\cdot 10^{-9}$ & $\left(\,0.21\pm0.42\pm0.84\,\right)\cdot 10^{-9}$\\
\hline
\end{tabular}
    \caption{Mean values of $\alpha_Y$ with 1$\sigma$ (68.3\%) and 2$\sigma$ (95.4\%) c.l.}
    \label{tab:Yukawa}
\end{table}

As one can see from this table, $\alpha_Y$ is quite small for both datasets, while SS+S2 data constrain $\alpha_Y$ around four times more than Solar System data alone. While still allowing for some GR deviation, the current analysis cannot discard GR.

\subsection{Buckingham gravity}
\begin{figure}[ht]
    \centering
    \includegraphics[width=.8\linewidth]{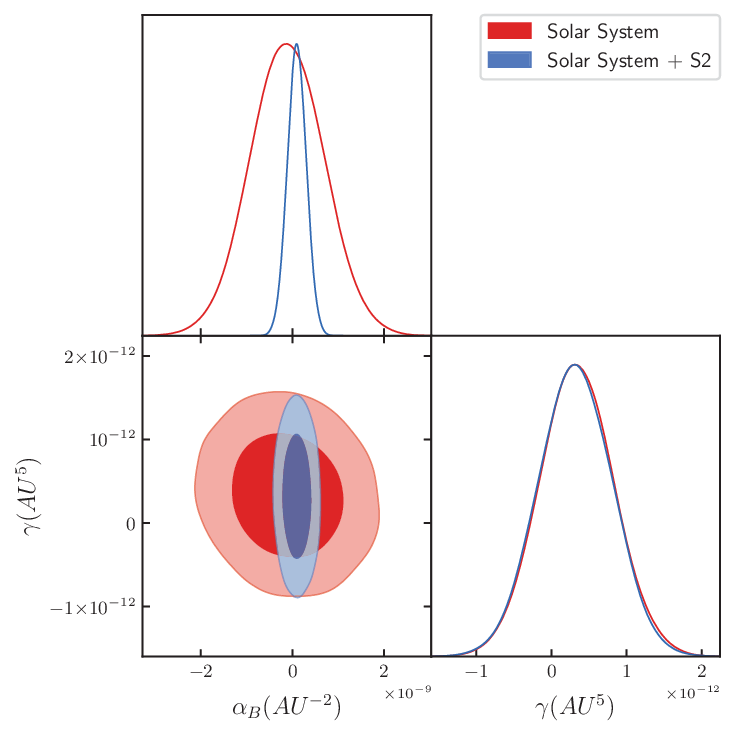}
    \caption{Posterior distributions for the $\alpha_B$ and $\gamma$ parameters in Buckingham gravity. In red, are the posteriors for the Solar System data alone. In blue, are the combined posteriors from Solar System+S2 data. Also shown are the 1$\sigma$ and 2$\sigma$ confidence contours of the joint analysis.}
    \label{fig:Buck}
\end{figure}

In the case of Buckingham gravity, the constraints over the parameters $\alpha_B$ and $\gamma$ can be seen in Fig. \ref{fig:Buck}.

As can be seen from this figure, similarly to Yukawa gravity, $\alpha_B$ is quite constrained when the S2 data are added  to the Solar System data. It also happens here that, from Solar System data alone, the best fit for $\alpha_B$ is negative, while when one adds the S2 constraint, we find $\alpha_B>0$ at the best fit. In the case of $\gamma$, however, there was no significant difference by adding the S2 data to the Solar System data. It should be expected because, as one can see from Eq. \eqref{eq:IV.14}, the term with $\gamma$ appears divided by $a^5$ and the  S2 star has a huge semimajor axis ($\sim10^3$ AU) when compared to Solar System planets. We can expect better constraints for $\gamma$ from smaller orbits.

Table \ref{tab:Buck} below shows the quantitative constraints.

\begin{table}[ht]
    \centering
    \begin{tabular} { l  c c}

 Parameter \hspace{1cm} & \hspace{1cm} SS Constraints \hspace{1cm} & \hspace{1cm} SS+S2 Constraints \hspace{1cm} \\
\hline
{\boldmath$\alpha_B$} (AU$^{-2}$) & $\left(\,-0.11\pm0.80\pm1.6\,\right)\cdot 10^{-9}$ & $\left(\,0.97\pm2.1\pm4.2\,\right)\cdot 10^{-10}$\\
{\boldmath$\gamma$} (AU$^5$) & $\left(\,0.34^{+0.49+1.0}_{-0.49-0.98}\,\right)\cdot 10^{-12}$ & $\left(\,0.32\pm0.49\pm0.97\,\right)\cdot 10^{-12}$\\
\hline
\end{tabular}
    \caption{Mean values of $\alpha_B$ and $\gamma$ with 1$\sigma$ (68.3\%) and 2$\sigma$ (95.4\%) c.l.}
    \label{tab:Buck}
\end{table}

As one can see from this table, $\alpha_B$ is quite small for both datasets, while SS+S2 data constrain $\alpha_B$ around four times more than Solar System data alone. As explained before, $\gamma$ is almost not affected by the addition of S2 to the Solar System data. Similarly to the Yukawa analysis, also here the data allow for some GR deviation, but GR cannot be discarded by the current analysis.

\section{Conclusions\protect\label{sec:Sec6}}

The Keplerian-type parametrization found in \citep{Benisty(2022)} for the solution of the two-body problem in Yukawa gravity was revisited.
This was done to solve some inconsistencies observed in the $\eta$ parametrization of true anomaly $\theta$. In particular, it was shown that the pericenter's advance must have the form $\delta\theta_{Yuk}=\zeta\left(e\right)\alpha m^{2}a^{2}$, where $\zeta\left(e\right)=\sqrt{1-e^{2}}/2$. This result is in agreement with Bertrand's theorem. In addition, we constructed a Kleplerian-type parametrization of the two-body problem in Buckingham gravity. This kind of gravity results from coupling the Buckingham potential (a modification of the Lennard-Jones intermolecular potential) with the gravitational Newtonian potential. The formulae for the advance of the pericenter derived as a byproduct of the Keplerian-type parametrization in both kinds of gravity were confirmed by using the Landau and Lifshitz's method. When we tested the models against Solar System and S2 precession data, we found that while some deviation from GR was allowed, GR could not be discarded by the current analysis. If more sources of data or more precise data are available in the future, the free parameters of the theories can be further constrained.

\begin{acknowledgments}
E. A. Gallegos would like to thank to Rodrigo R. Cuzinatto for useful discussions about this work. R. Perca Gonzales would like to thank to Elmer Luque Canaza for useful discussions about observational tests of modified gravity and acknowledge financial support from the Universidad Nacional de San Agust\'in de Arequipa (Grant No IBA-IB-35-2020-UNSA). J. F. Jesus acknowledges financial support from  {Conselho Nacional de Desenvolvimento Cient\'ifico e Tecnol\'ogico} (CNPq) (No. 314028/2023-4). The authors would also like to thank the anonymous reviewer for their invaluable comments and suggestions.
\end{acknowledgments}

\appendix

\section{Landau and Lifshitz 's method for computing the pericenter precession
\protect\label{sec:AppendixA}}

Here we verify our results for the advance of the pericenter found in the two-body problem both in Yukawa gravity and in Buckingham gravity.
This is done by using the method described in \citep{LandauLifshitz(1976)} and that is based upon (\ref{eq:II.3}), i. e., the expression that
gives the angular displacement $\Delta\theta$ of the ``body'' when this moves from $r_{min}$ to $r_{max}$ and then returns to $r_{min}$.
To avoid spurious divergences, one rewrites (\ref{eq:II.3}) in the alternate form
\begin{equation}
\Delta\theta=-2\frac{\partial}{\partial l}\int_{r_{min}}^{r_{max}}\sqrt{2\left(\epsilon-v\left(r\right)\right)-l^{2}/r^{2}}\,dr-2\int_{r_{min}}^{r_{max}}\frac{\partial v/\partial l\,dr}{\sqrt{2\left(\epsilon-v\left(r\right)\right)-l^{2}/r^{2}}}.\label{eq:A1}
\end{equation}
Note that in the derivation of (\ref{eq:A1}) we are considering the possibility of the potential $v\left(r\right)$ to depend on the angular
momentum per reduced mass, $l$. In fact, this happens when one considers the relativistic correction: $v_{GR}=-\kappa l^{2}/\left(c^{2}r^{3}\right)$.

To proceed, we shall regard the Yukawa potential, the Buckingham potential and the relativistic contribution as small corrections, $\delta v$,
to the Newtonian gravitational potential $v_{grav}=-\kappa/r$. Hence, substituting $v=-\kappa/r+\delta v$ in (\ref{eq:A1}) and expanding
the integrands up to first order in $\delta v$, one obtains
\begin{eqnarray}
\Delta\theta & = & -2\frac{\partial}{\partial l}\int_{r_{min}}^{r_{max}}\sqrt{2\left(\epsilon+\kappa/r\right)-l^{2}/r^{2}}\,dr+2\frac{\partial}{\partial l}\int_{r_{min}}^{r_{max}}\frac{\delta v\,dr}{\sqrt{2\left(\epsilon+\kappa/r\right)-l^{2}/r^{2}}}\nonumber \\
 &  & -2\int_{r_{min}}^{r_{max}}\frac{\partial\delta v/\partial l\,dr}{\sqrt{2\left(\epsilon+\kappa/r\right)-l^{2}/r^{2}}}\label{eq:A2}
\end{eqnarray}
It is easy to verify that the zero-order term in (\ref{eq:A2}) is equal to $2\pi$, a result that is in accordance with Bertrand's theorem.
On the other hand, the last two terms, i.e., the first-order terms in $\delta v$, give the displacement of the pericenter at this order
of approximation. After changing the variable of integration from $r$ to $\theta$ by means of the ``unperturbed'' elliptical equation
of motion (\ref{eq:II.4}), one can see that the displacement of the pericenter $\delta\theta$ is given by
\begin{equation}
\delta\theta=\frac{\partial}{\partial l}\left(\frac{2}{l}\int_{0}^{\pi}r^{2}\delta v\,d\theta\right)-\frac{2}{l}\int_{0}^{\pi}r^{2}\frac{\partial\delta v}{\partial l}\,d\theta.\label{eq:A3}
\end{equation}
This formula is the natural extension of that derived in \citep{LandauLifshitz(1976)}, considering the possibility of the perturbation $\delta v$ to depend on the angular momentum $l$. It is important to stress that before taking the $l$ derivatives indicated in (\ref{eq:A3}), the semilatus rectum $p$ and the eccentricity $e$ of the ``unperturbed'' orbit must be expressed in terms of the angular momentum $l$, namely, $p=l^{2}/\kappa$ and $e=\sqrt{1-l^{2}/\kappa a}$.

Focusing our attention, first, in the potential $v\left(r\right)$ defined in Yukawa gravity [see Eq. (\ref{eq:III.1})], we easily identify
the perturbation $\delta v_{Yuk-GR}$ as
\begin{eqnarray}
\delta v_{Yuk-GR} & = & -\frac{\kappa\alpha}{r}\text{e}^{-mr}-\frac{\kappa l^{2}}{c^{2}r^{3}}\nonumber \\
 & = & -\frac{\kappa\alpha}{r}\left(1-mr+\frac{m^{2}r^{2}}{2}\right)-\frac{\kappa l^{2}}{c^{2}r^{3}},\label{eq:A4}
\end{eqnarray}
where in the last equality the exponential expansion was truncated up to second order in $mr$. 

Substituting (\ref{eq:A4}) into (\ref{eq:A3}) and performing the angular integrals, after eliminating $r$ in terms of $\theta$ through
the elliptical equation, $r=p/\left(1+e\text{cos}\theta\right)$, and the $l-$derivative, one obtains
\begin{equation}
\delta\theta_{Yuk+GR}=\pi\sqrt{1-e^{2}}f+6\pi\beta,\label{eq:A5}
\end{equation}
where, as before, $f\equiv\alpha m^{2}a^{2}$ and $\beta\equiv\kappa/\left[ac^{2}\left(1-e^{2}\right)\right]$.
This result for the advance of the pericenter is in agreement with that found in (\ref{eq:III.14}), after ignoring the $f\beta$  coupling
terms and dividing by $2\pi$. 

Finally, the perturbation $\delta v_{Buck-GR}$ in Buckingham gravity [see Eq. (\ref{eq:IV.1})] reads
\begin{eqnarray}
\delta v_{Buck-GR} & = & \kappa\alpha\text{e}^{-mr}-\frac{\kappa\gamma}{r^{6}}-\frac{\kappa l^{2}}{c^{2}r^{3}}\nonumber \\
 & = & \kappa\alpha\left(1-mr\right)-\frac{\kappa\gamma}{r^{6}}-\frac{\kappa l^{2}}{c^{2}r^{3}}.\label{eq:A6}
\end{eqnarray}
Substituting $\delta v_{Buck-GR}$ into (\ref{eq:A3}) and performing the angular integrals and the $l$ derivative, one finds
\begin{equation}
\delta\theta_{Buck-GR}=2\pi\sqrt{1-e^{2}}\tilde{f}+\frac{15\pi\left(8+12e^{2}+e^{4}\right)\gamma/a^{5}}{4\left(1-e^{2}\right)^{5}}+6\pi\beta,
\end{equation}
where, as before, $\tilde{f}=\alpha ma^{2}$. This result for the advance of the pericenter in the two-body problem in Buckingham gravity is
in accordance with that found in (\ref{eq:IV.14}), after ignoring the $\tilde{f}\beta$ and $\gamma\beta$ coupling terms and dividing by $2\pi$.

\section{Functions defined in Buckingham gravity\protect\label{sec:AppendixB}}

{\small The functions $F\left(e,\rho\right)$ and $G\left(e,\rho\right)$
that appear in (\ref{eq:IV.5}) are given by}
\begin{eqnarray}
F\left(e,\rho\right) & = & 42+\left(3e^{6}-15e^{4}+27e^{2}-25\right)e^{2}+4\left(3e^{4}-8e^{2}-27\right)e\rho+3\left(e^{6}-11e^{4}+\right.\nonumber \\
 &  & \left.+35e^{2}+39\right)e^{2}\rho^{2}+4\left(3e^{4}-20e^{2}-15\right)e^{3}\rho^{3}+4\left(5e^{2}+3\right)e^{4}\rho^{4}\label{eq:AppB.1}
\end{eqnarray}
\begin{eqnarray}
G\left(e,\rho\right) & = & -174+\left(3e^{8}-28e^{6}+92e^{4}-134e^{2}+177\right)e^{2}+2\left(5e^{8}-19e^{6}-17e^{4}-93e^{2}+\right.\nonumber \\
 &  & \left.+284\right)e\rho+\left(3e^{8}-46e^{6}+368e^{4}-178e^{2}-787\right)e^{2}\rho^{2}+2\left(11e^{6}-161e^{4}+\right.\nonumber \\
 &  & \left.+185e^{2}+285\right)e^{3}\rho^{3}+4\left(25e^{4}-52e^{2}-53\right)e^{4}\rho^{4}-8\left(e^{4}-5e^{2}-4\right)e^{5}\rho^{5},\label{eq:AppB.2}
\end{eqnarray}
where $\rho=\cos\eta$.

The functions $F_{n}\left(e,\beta\right)$ that appear in the Kepler-type equation (\ref{eq:IV.7}) are given by
\begin{eqnarray}
F_{1}\left(e,\beta\right) & = & 64+27e^{2}+21e^{4}+\beta\left(288-17e^{2}+159e^{4}+18e^{6}\right)+3\left(4+3e^{2}\right)\xi\left(e,\beta\right)\label{eq:AppB.3}
\end{eqnarray}
\begin{eqnarray}
-F_{2}\left(e,\beta\right)/2 & =39+ & 12e^{2}+5e^{4}+\beta\left(141+30e^{2}+47e^{4}+6e^{6}\right)+\frac{3}{2}\left(6+e^{2}\right)\xi\left(e,\beta\right)\label{eq:AppB.4}
\end{eqnarray}
\begin{equation}
F_{3}\left(e,\beta\right)=31+17e^{2}+\beta\left(83+83e^{2}+26e^{4}\right)+9\xi\left(e,\beta\right)\label{eq:AppB.5}
\end{equation}
\begin{equation}
-F_{4}\left(e,\beta\right)/2=2+2e^{2}+\beta\left(3+10e^{2}+3e^{4}\right)+\frac{3}{4}\xi\left(e,\beta\right),\label{eq:AppB.6}
\end{equation}
where $\xi\left(e,\beta\right)\equiv\sqrt{1-e^{2}}\left[4+e^{2}+\beta\left(16+18e^{2}+e^{4}\right)\right]$.

The $e-$polynomials $P_{n}\left(e\right)$ in Eqs. (\ref{eq:IV.10}) and (\ref{eq:IV.11}) are given by
\begin{equation}
P_{0}\left(e\right)=5\left(24+28e^{2}+63e^{4}-6e^{6}+3e^{8}\right);\,\,\,\,\,\,\,P_{1}\left(e\right)=-16\left(20+21e^{2}+15e^{4}\right)\label{eq:AppB.7}
\end{equation}
\begin{equation}
P_{2}\left(e\right)=4\left(45+60e^{2}+5e^{4}+2e^{6}\right);\,\,P_{3}\left(e\right)=-16\left(3+5e^{2}\right);\,\,P_{4}\left(e\right)=5+10e^{2}+e^{4}.\label{eq:AppB.8}
\end{equation}

Finally, the functions $\tilde{P}_{n}\left(e,\beta\right)$ in (\ref{eq:IV.12}) are defined as:
\begin{eqnarray}
-\tilde{P}_{1}\left(e,\beta\right)/2 & = & 1280+1752e^{2}+860e^{4}+35e^{6}+\beta\left(5760+7096e^{2}+\right.\nonumber \\
 &  & \left.4788e^{4}-725e^{6}+70e^{8}\right)\label{eq:AppB.9}
\end{eqnarray}
\begin{eqnarray}
\tilde{P}_{2}\left(e,\beta\right)/4 & = & 1160+820e^{2}+275e^{4}-11e^{6}+\beta\left(4920+3892e^{2}+\right.\nonumber \\
 &  & \left.+1115e^{4}-230e^{6}+11e^{8}\right)\label{eq:AppB.10}
\end{eqnarray}
\begin{eqnarray}
-\tilde{P}_{3}\left(e,\beta\right) & = & 3184+1690e^{2}+175e^{4}-\beta\left(-12912-9386e^{2}+\right.\nonumber \\
 &  & \left.+451e^{4}+4e^{6}\right)\label{eq:AppB.11}
\end{eqnarray}
\begin{eqnarray}
\tilde{P}_{4}\left(e,\beta\right)/2 & = & 490+265e^{2}-7e^{4}+\beta\left(1914+1547e^{2}+\right.\nonumber \\
 &  & \left.-232e^{4}+7e^{6}\right)\label{eq:AppB.12}
\end{eqnarray}
\begin{equation}
-\tilde{P}_{5}\left(e,\beta\right)=114+73e^{2}+\frac{\beta}{5}\left(2154+2067e^{2}-176e^{4}\right).\label{eq:AppB.13}
\end{equation}

\end{document}